\documentclass[journal=nalefd,layout=twocolumn,manuscript=letter]{achemso}

\usepackage{newtxtext,newtxmath} 

\usepackage{caption}
\usepackage{setspace}
\usepackage{titlesec}
\titlespacing*{\section}{0pt}{1ex}{0.5ex}
\titlespacing*{\subsection}{0pt}{0.8ex}{0.4ex}

\usepackage{chemformula} 
\usepackage[T1]{fontenc} 
\usepackage{enumitem}    
\usepackage{graphicx}    
\usepackage{braket}      
\usepackage{xcolor}      
\usepackage{amsmath}     

\makeatletter
\newlength{\@myfigwidth}
\if@twocolumn
\else
\fi
\newcommand{\myfigwidth}{\@myfigwidth}
\makeatother

\author{Patrick Grenzer}
\affiliation[JMU]
{Institute of Physical and Theoretical Chemistry, Julius-Maximilians-University Würzburg, Germany}
\author{Fabian Lie}
\affiliation[Twente]
{MESA+ Institute for Nanotechnology, University of Twente, Enschede, The Netherlands}
\author{Klaus H. Eckstein}
\affiliation[JMU]
{Institute of Physical and Theoretical Chemistry, Julius-Maximilians-University Würzburg, Germany}
\author{Tobias Hertel}
\affiliation[JMU]
{Institute of Physical and Theoretical Chemistry, Julius-Maximilians-University Würzburg, Germany}
\email{tobias.hertel@uni-wuerzburg.de}
\author{Linn Leppert}
\affiliation[Birmingham]
{School of Metallurgy and Materials, University of Birmingham, Birmingham, United Kingdom}
\alsoaffiliation[Twente]
{MESA+ Institute for Nanotechnology, University of Twente, Enschede, The Netherlands}
\email{l.leppert@bham.ac.uk}

\title[]{Intra- and Interlayer Excitonic Finestructure of the Two-Dimensional Perovskite (PEA)$_2$PbI$_4$}

\begin{document}


\begin{abstract}
Two-dimensional halide perovskites host strongly bound excitons whose fine structure controls polarization selection rules and radiative recombination, yet several spectral features in \((\mathrm{PEA})_2\mathrm{PbI}_4\) remain controversially assigned. Here, polarization-resolved low-temperature photoluminescence combined with first-principles \(G_0W_0\)+BSE calculations resolves both the intralayer and interlayer excitonic fine structure of this prototypical \(n=1\) Ruddlesden-Popper perovskite. The low-energy multiplet is consistently described as a purely excitonic intralayer fine structure governed by crystal symmetry, octahedral distortions, and the two-layer unit cell, without invoking Rashba or exciton-polaron mechanisms as the primary origin. A weaker doublet \(\sim 45\) meV above the bright intralayer states is identified as interlayer excitons from its agreement with the calculated interlayer manifold in energy and splitting. Although the static calculations underestimate their oscillator strength and do not reproduce the observed orthogonal polarizations, distortion-induced mixing with bright intralayer excitons strongly enhances interlayer optical activity and provides a plausible explanation for their visibility. Our results establish interlayer excitons in \((\mathrm{PEA})_2\mathrm{PbI}_4\) and refine the excitonic description of fine structure in two-dimensional perovskites.
\end{abstract}

\section{Introduction}
Two-dimensional (2D) lead-halide perovskites are layered semiconductors in which inorganic sheets are separated by bulky organic cations thus forming natural quantum wells.\cite{Blancon20} In the monolayer limit ($n=1$), exemplified here 
by phenethylammonium lead iodide ((PEA)$_2$PbI$_4$), quantum and dielectric confinement are particularly strong, giving rise to tightly bound excitons with binding energies of several hundred meV and pronounced excitonic effects even at room temperature.\cite{Tanaka05,Shimizu06,Straus16} These properties make $n=1$ Ruddlesden-Popper perovskites a useful platform for studying how reduced dimensionality, lattice symmetry, and spin-orbit coupling affect excitonic states and their optical selection rules.

A central feature of these materials is their excitonic fine structure, which controls polarization selection rules, radiative recombination pathways, and the low-temperature photoluminescence (PL) spectrum. In (PEA)$_2$PbI$_4$, the commonly used fine-structure picture is described as starting from electron-hole exchange separating a predominantly dark singlet state \(\mathrm{D}\), conventionally associated with total angular momentum $J=0$, from a bright triplet with $J=1$.\cite{Ema06,Neutzner18,Sercel19,Fang20,Dyksik21b,Baranowski22b} Within the bright manifold, reduced crystal symmetry distinguishes the in-plane bright states \(\mathrm{X}\) and \(\mathrm{Y}\) from the out-of-plane bright state \(\mathrm{Z}\),\cite{Posmyk23,Posmyk24b} while octahedral distortions lift the near-degeneracy of the in-plane states and define their approximately orthogonal polarization axes.\cite{Posmyk23,Baranowski22b,becker18,Sercel19,Weinberg23} 

Experimentally, low-temperature spectroscopy has established that (PEA)$_2$PbI$_4$ exhibits a nontrivial excitonic multiplet.\cite{Fang20,Do20b,CanetAlbiach22,Posmyk22,Posmyk23,Zhang23} Time-resolved PL identified a dark-bright splitting of about 10\,meV,\cite{Fang20} while high-field polarization spectroscopy resolved in-plane splittings on the meV scale.\cite{Do20b} Magneto-optical studies across related compounds further mapped field-dependent level shifts, bright-dark mixing, and exciton-phonon coupling.\cite{Dyksik20,Dyksik21,Dyksik21b,Dyksik24,Posmyk22,Posmyk23,Posmyk24} The existence of a low-energy fine structure is therefore well established experimentally. By contrast, its microscopic interpretation remains debated, and the origin of weak higher-energy spectral features has been even less clear.

To account for the observed low-energy PL structure, previous discussions have invoked crystal-field effects, Rashba-type splittings, exciton-polaron sidebands, and other symmetry-related excitonic descriptions.\cite{Sercel23,Sercel19,Biffi23,MolinaSanchez18,Zuri24} More recent first-principles and model Hamiltonian studies have clarified the effect of exchange, spin-orbit coupling, symmetry lowering, and structural distortions on the intralayer exciton manifold in layered halide perovskites.\cite{Filip22,Quarti24,Sercel19,Biffi23,MolinaSanchez18} At the same time, these studies indicate that in compounds such as (PEA)$_2$PbI$_4$, the presence of two inorganic layers in the crystallographic unit cell introduces an additional layer degree of freedom and can therefore enlarge the excitonic manifold beyond the conventional $\mathrm{D/X/Y/Z}$ picture.\cite{Filip22,Quarti24}

Despite this progress, two points have remained open. First, it is still unclear whether the low-energy PL multiplet requires mechanisms such as Rashba splitting or exciton-polaron sidebands as its primary explanation, or whether it can be understood within a symmetry- and layer-resolved excitonic framework. Second, the microscopic character of the weak higher-energy doublet has remained unresolved.

In this Letter, we combine temperature- and polarization-resolved PL spectroscopy of mechanically exfoliated (PEA)$_2$PbI$_4$ crystals with first-principles \(G_0W_0\)+BSE calculations to address these open questions. Comparison between experiment and theory shows that the low-energy bright multiplet is captured by a symmetry- and layer-resolved excitonic picture and does not require Rashba splitting or exciton-polaron sidebands as its primary explanation. The same comparison identifies the weak higher-energy doublet as interlayer excitons. Remaining discrepancies in optical visibility and polarization indicate additional lattice-coupled mixing beyond the static ground-state description.

\section{Results and Discussion}

\begin{figure}[htbp]
	\centering
		\includegraphics[width=0.9\columnwidth]{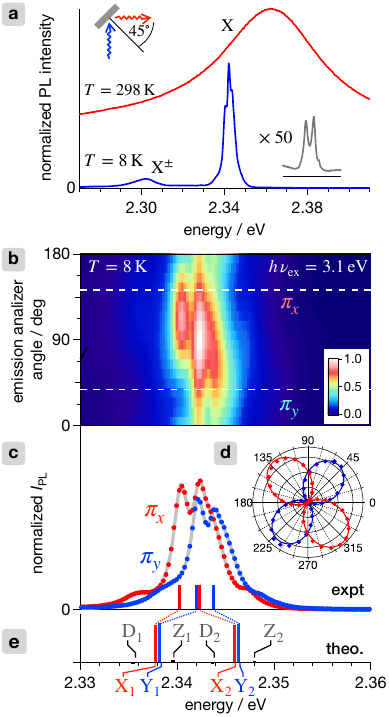}
		\caption{(a) Room- and low-temperature PL spectra. Inset: magnified view of the weak higher-energy doublet from a representative low-temperature spectrum of a comparable flake. (b) Polarization-resolved PL map of the intralayer exciton manifold. (c,d) Decomposition into two polarized spectral components, $\pi_x$ and $\pi_y$, and their polar representations. (e) Comparison with the calculated exciton energies and oscillator strengths.}
		\label{fig1}
\end{figure}

Figure~\ref{fig1}a shows PL spectra of a thin, mechanically exfoliated $\mathrm{(PEA)_2PbI_4}$ flake recorded at room temperature and at $8\,\mathrm{K}$. The excitation with $p$-polarized light and the detection geometry are indicated schematically in the inset. As reported previously, the room-temperature spectrum is largely featureless because of thermal broadening, whereas at low temperature three distinct emission regions appear near 2.30, 2.34, and $2.38\,\mathrm{eV}$.\cite{Do20b,Neutzner18} The weak low-energy feature can be attributed to a charged exciton (trion, \(\mathrm{X}^\pm\)), consistent with residual doping in these samples.\cite{Ziegler23} The central band is assigned to the fine structure of the bright intralayer excitons. The weak higher-energy doublet is addressed separately below.

We first examine the polarization dependence of the dominant exciton emission between about 2.332 and $2.354\,\mathrm{eV}$ (Fig.~\ref{fig1}b). The spectra are dominated by two nearly orthogonal linear-polarization components, denoted \(\pi_x\) and \(\pi_y\), shown in Fig.~\ref{fig1}c, with their polarization dependence displayed in Fig.~\ref{fig1}d. Although these spectra reveal structured bright multiplets, polarization maps alone do not unambiguously assign the observed lines to the predicted \(\mathrm{D}\), \(\mathrm{X}\), \(\mathrm{Y}\), and \(\mathrm{Z}\) excitonic manifold. We therefore compare the measured ordering and polarization pattern with a microscopic excitonic model.

To assign this experimentally observed ordering and polarization pattern, we compare the experiment with first-principles \(G_0W_0\)+BSE calculations of the excitonic states for a fully relaxed \(\mathrm{(PEA)_2PbI_4}\) structure\cite{Biega23} including spin-orbit coupling, performed using \textsc{berkeleygw}.\cite{Deslippe12,DelBen19,Barker22} Computational details and convergence tests are provided in the SI. In the following, we refer to this structure as the relaxed reference structure.

\begin{figure*}[htbp]
	\centering
		\includegraphics[width=0.9\textwidth]{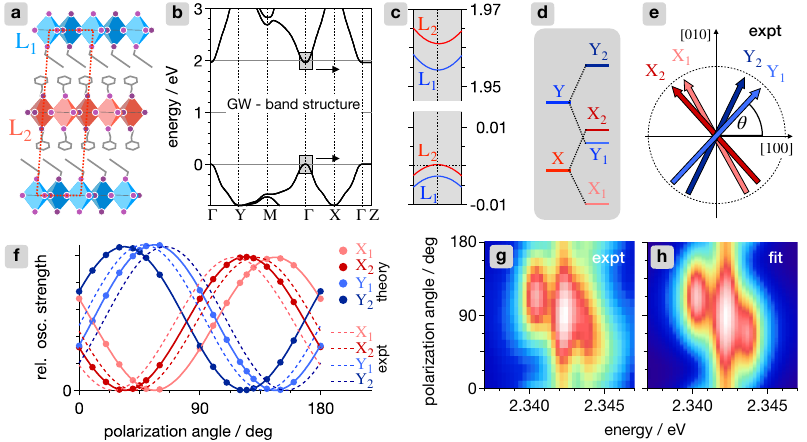}
        \caption{(a) Crystal structure with two symmetry-inequivalent layers, \(\mathrm{L}_1\) and \(\mathrm{L}_2\), in the unit cell, viewed along the \(a\) axis of the triclinic unit cell. (b,c) Calculated band structure and expanded view of the valence- and conduction-band edges relevant to the exciton fine structure. (d) Schematic illustration of the resulting \(\mathrm{L_1/L_2}\)-resolved doubling of the dominant bright transitions \(\mathrm{X}_1\), \(\mathrm{X}_2\), \(\mathrm{Y}_1\), and \(\mathrm{Y}_2\). (e) Calculated transition-dipole orientations of the four dominant fine-structure transitions and definition of the polarization angle. (f) Calculated and experimentally fitted polarization dependencies of the four dominant excitonic transitions. (g,h) Experimental PL polarization map and corresponding fit, showing the correspondence underlying the assignment.}
		\label{fig2}
\end{figure*}

Because one-shot \(G_0W_0\)+BSE calculations starting from semilocal density functional theory (DFT) commonly underestimate absolute exciton transition energies in halide perovskites,\cite{Filip14d,Brivio14,Scherpelz16,Wiktor17a,Leppert19,Filip22,Ohad22} we align theory and experiment by applying a rigid blue shift of \(0.491\,\mathrm{eV}\) to the calculated intralayer manifold. This alignment is used only to compare the relative energetic ordering and polarization character of the exciton transitions. As shown by the stick spectra in Fig.~\ref{fig1}e, the calculated bright intralayer sequence reproduces the observed alternation between the dominant lines between the \(\pi_x\) and \(\pi_y\) polarization channels. This agreement in relative ordering and polarization character provides the basis for assigning the experimentally resolved bright transitions.
The two-layer unit cell of \(\mathrm{(PEA)_2PbI_4}\) shown in Fig.~\ref{fig2}a introduces a layer degree of freedom that doubles the intralayer exciton manifold. This is reflected in the doubling of the valence- and conduction-band states in the calculated \(GW\) band structure in Fig.~\ref{fig2}b and in the corresponding layer-dependent offsets of the band edges shown in the close-up in Fig.~\ref{fig2}c. The resulting ordering of the experimentally observed bright \(\mathrm{X}\) and \(\mathrm{Y}\) doublets is illustrated schematically in Fig.~\ref{fig2}d. The predicted internal ordering within each layer-resolved intralayer multiplet, \(\mathrm{(D_L,X_L,Y_L,Z_L)}\) with \(\mathrm{L}\in\{1,2\}\), is consistent with recent work on exciton fine structure in layered lead-halide perovskites.\cite{Posmyk22,Quarti24} What is added here is the explicit duplication into two similar \(\mathrm{L}_1\) and \(\mathrm{L}_2\) manifolds arising from the two-layer unit cell of \(\mathrm{(PEA)_2PbI_4}\).  The internal spacing within each multiplet are nearly identical, so that the calculated \(\mathrm{L}_1\) and \(\mathrm{L}_2\) manifolds differ mainly by a layer-dependent offset.

The main quantitative discrepancy between theory and experiment is that the calculated dominant bright transitions \(\mathrm{X}_1\), \(\mathrm{Y}_1\), \(\mathrm{X}_2\), and \(\mathrm{Y}_2\) exhibit a smaller \(\mathrm{X}\)–\(\mathrm{Y}\) centroid spacing, such that the \(\mathrm{L}_1\) and \(\mathrm{L}_2\) manifolds overlap in the measured spectra rather than remaining clearly separated as in theory. The corresponding \(\mathrm{D}\) and \(\mathrm{Z}\) partners are predicted to bracket each bright \(\mathrm{X}\)–\(\mathrm{Y}\) pair, but their experimental identification is less direct and is therefore treated more cautiously below. A complete summary of calculated energies, oscillator strengths, and dipole characters is provided in the SI.

Using the calculated ordering and polarization character as a guide for the line assignment, we globally fit the polarization-resolved PL data in Fig.~\ref{fig1}b using \(\cos^2\)-type angular dependencies and variable transition energies. The resulting fit is shown in Fig.~\ref{fig2}h for direct comparison alongside the experimental data in Fig.~\ref{fig2}g. From this analysis, we extract four dominant bright transitions at \(2.3402\)\,eV (\(\mathrm{X}_1\)), \(2.3420\)\,eV (\(\mathrm{Y}_1\)), \(2.3423\)\,eV (\(\mathrm{X}_2\)), and \(2.3438\)\,eV (\(\mathrm{Y}_2\)), each with an uncertainty of \(\pm 0.2\)\,meV. These four lines resolve into two closely spaced \(\mathrm{X/Y}\)-like pairs, consistent with a bright intralayer manifold split by the two-layer unit cell into \(\mathrm{L}_1\)- and \(\mathrm{L}_2\)-resolved components. The corresponding polarization angles are summarized in Fig.~\ref{fig2}e. For comparison with theory, Fig.~\ref{fig2}f shows the calculated angular dependencies of the four bright transitions together with the corresponding experimental \(\cos^2\)-type fit functions.

These four lines define the most robust experimental energy scales of the bright manifold, namely an \(\mathrm{L_1/L_2}\) splitting of \(2.0\)\,meV between the two bright pairs and an intralayer \(\mathrm{Y\!-\!X}\) anisotropy of about \(1.5\)–\(1.8\)\,meV within each pair. Additional weak shoulders tentatively suggest the location of the corresponding \(\mathrm{D}\) and \(\mathrm{Z}\) transitions. These weaker features are therefore used more cautiously below and serve only as an exploratory basis for estimating broader \(\mathrm{D/X/Y/Z}\) centroids and additional splittings.

Against this experimental baseline, the calculation for the relaxed reference structure captures the qualitative \(\mathrm{L_1/L_2}\)-resolved organization and polarization character of the bright intralayer manifold, but not all quantitative details. The assignment therefore rests primarily on the correct relative ordering and polarization character of the dominant bright lines, rather than on quantitative agreement in the extracted splittings. In particular, calculations overestimate the \(\mathrm{L_1/L_2}\) splitting of the two bright pairs, predicting \(\sim 8\)\,meV rather than the experimental \(1.5\)–\(1.8\)\,meV, while underestimating the intralayer bright-state anisotropy, yielding only \(\sim 0.4\)\,meV instead of the observed \(2.0\)\,meV. The calculated transition dipoles also reproduce the predominantly in-plane character of the bright states. Together with the correct relative ordering of the dominant bright lines, this provides the basis for assigning the manifold and identifying its \(\mathrm{L_1/L_2}\)-resolved excitonic character. Residual discrepancies remain in the relative splittings and in the detailed polarization angles.

To clarify the origin of the overestimated \(\mathrm{L_1/L_2}\) splitting in the relaxed reference structure, we compared \(\mathrm{(PEA)_2PbI_4}\) with two auxiliary Cs-based models (see SI for structural definitions, tabulated excitation energies, and additional discussion). In the first model, the organic PEA spacers were replaced by Cs ions while retaining the distorted inorganic framework of the reference structure. In the second, idealized \(\mathrm{Cs_2PbI_4}\) model, both the organic spacers and the octahedral distortions were removed, restoring a highly symmetric, untilted structure.

Comparing these three structures shows that the substantial layer-dependent offset between the two bright manifolds is unique to the fully relaxed PEA structure. In the idealized \(\mathrm{Cs_2PbI_4}\) limit, the \(\mathrm{L_1/L_2}\) splitting of the bright manifolds collapses to nearly zero. In the Cs-substituted model that retains the distorted inorganic framework, a small residual \(\mathrm{X\!-\!Y}\) anisotropy remains, but the layer-dependent offset is negligible. Only the full low-symmetry PEA environment yields the pronounced calculated \(\sim 8\)\,meV \(\mathrm{L_1/L_2}\) splitting. This indicates that the calculated \(\mathrm{L_1/L_2}\) splitting is driven primarily by a layer-dependent offset arising from the inequivalence of the organic dielectric environment surrounding the two inorganic layers in the relaxed structure, rather than by strong hybridization between the corresponding layer-resolved excitonic configurations. The low-energy fine structure is therefore described within a symmetry- and layer-resolved excitonic picture. 

Since the relaxed reference structure is centrosymmetric, the present calculations do not exhibit Rashba-type band-edge splitting. Moreover, the calculations up to this point do not include exciton-phonon coupling effects. The qualitative agreement with experiment in the relative ordering and polarization character of the dominant bright lines therefore suggests that previously proposed mechanisms related to Rashba splitting or exciton-polaron sidebands are not required as the primary explanation for the observed fine structure.

\begin{figure}[ht]
	\centering
		\includegraphics[width=\columnwidth]{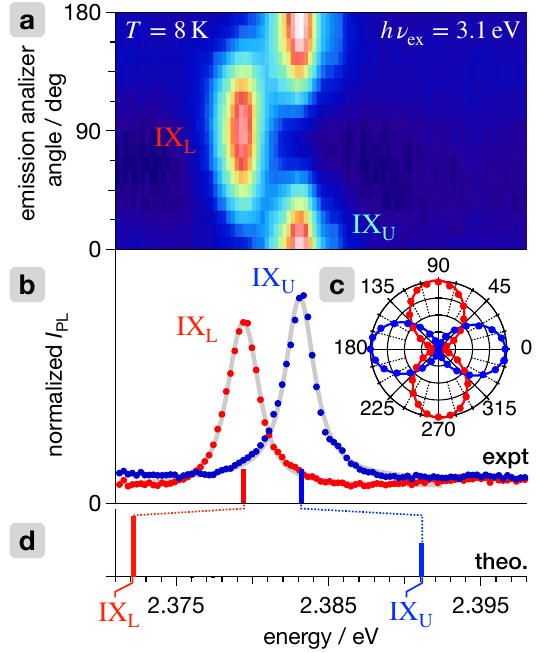}
		\caption{Polarization-resolved PL of the higher-energy exciton doublet at 8\,K. (a) Experimental PL intensity map. (b) Polarization-selected spectra of the lower and upper components of the higher-energy doublet, labeled \(\mathrm{IX}_L\) and \(\mathrm{IX}_U\). (c) Normalized polar intensity representation. (d) Calculated interlayer exciton transition energies.}
		\label{fig3}
\end{figure}

\begin{figure*}[ht]
	\centering
		\includegraphics[width=0.9\textwidth]{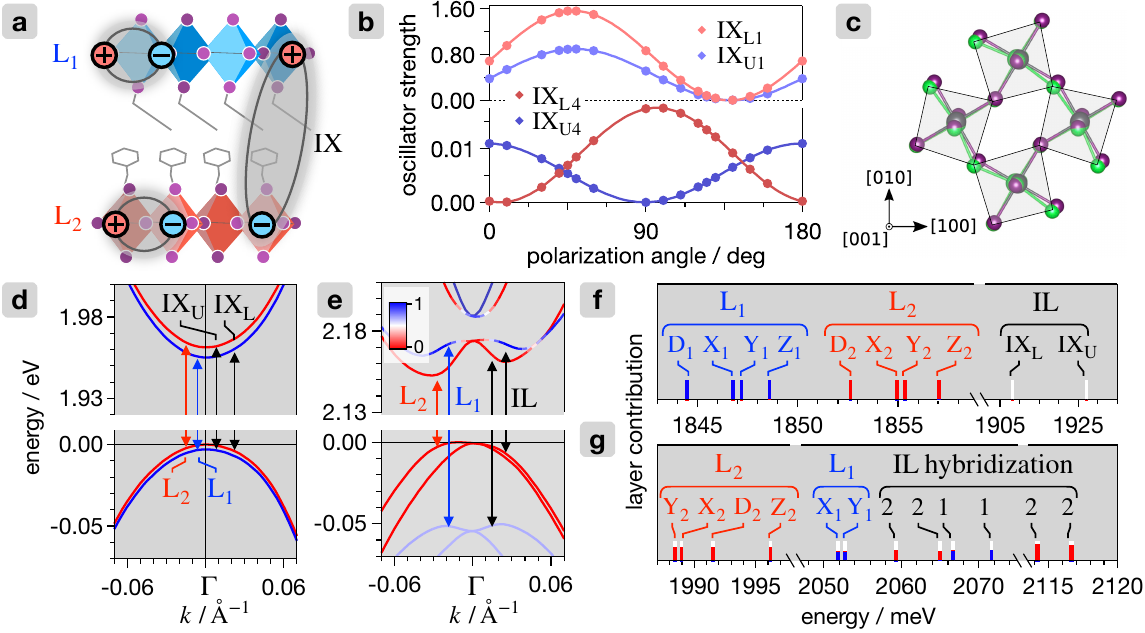}
		\caption{Interlayer-exciton assignment and distortion-based interpretation of its optical activity. (a) Schematic illustration of intra- and interlayer excitons in the two-layer unit cell. (b) Calculated polarization dependencies of selected interlayer bands. (c) Schematic phonon-induced lattice distortion. (d) Valence- and conduction-band edges in the reference structure, with intra- and interlayer transitions indicated schematically. (e) Band structure of the phonon-displaced model, with \(\mathrm{L}_2\) layer projection shown by false color. (f) Layer contributions of the excitonic states in the reference structure, identifying the relevant weak higher-energy states as interlayer in character. (g) Excitonic states of the phonon-displaced model. The shifted \(L_2\) manifold and \(X_1/Y_1\) remain identifiable, whereas the higher-energy states grouped under ``IL hybridization'' show substantial intra-/interlayer mixing. Numbers indicate degeneracies.}
		\label{fig4}
\end{figure*}

As noted at the beginning of this discussion, the PL spectrum also contains a distinct weak doublet at \(2.3794\pm0.0002\)\,eV and \(2.3832\pm0.0002\)\,eV, about 45\,meV above the onset of the intralayer exciton manifold. We now examine this higher-energy feature separately, because it raises a different assignment problem. Polarization-resolved PL at 8\,K shows the two components, labeled \(\mathrm{IX}_L\) and \(\mathrm{IX}_U\), to have nearly orthogonal in-plane polarization axes (Fig.~\ref{fig3}). Their much lower intensity than the dominant intralayer bands is consistent with a weakly allowed excitonic manifold.

\(G_0W_0\)+BSE calculations using the relaxed reference structure predict two weak four-fold degenerate exciton sets in about the same general spectral range, lying about 58\,meV above the bright intralayer excitons, compared with about 45\,meV in experiment, and exhibiting a larger internal splitting of 19\,meV versus 4\,meV in experiment. The computed energies of these interlayer excitons are shown in Fig.~\ref{fig3}d after applying an independent rigid shift of \(+0.464\,\mathrm{eV}\) to align the interlayer manifold with experiment, indicating a modest quantitative mismatch in the interlayer--intralayer separation. Their real-space probability distributions show clear interlayer character, with electron and hole residing on adjacent inorganic layers (see the schematic illustration in Fig.~\ref{fig4}a). The lower-energy set is dominated by \(\mathrm{L_2\!\rightarrow L_1}\) transitions between the inner band-edge states defined by the layer-dependent offset (see Fig.~\ref{fig2}c and also Fig.~\ref{fig4}d), whereas the higher-energy set is dominated by \(\mathrm{L_1\!\rightarrow L_2}\) transitions between the outer band-edge states. This supports assigning the observed PL doublet to the corresponding interlayer exciton manifold.

Despite the reasonable agreement in energetic placement and manifold structure, the calculated optical activity of the interlayer states does not fully account for the experimental observations. The two calculated interlayer excitons with the largest oscillator strengths in Fig.~\ref{fig4}b have nearly parallel in-plane dipole orientations, whereas the observed \(\mathrm{IX}_L\) and \(\mathrm{IX}_U\) peaks are nearly orthogonally polarized. A second pair of calculated interlayer states, corresponding to \(\mathrm{L_2\!\rightarrow L_1}\) and \(\mathrm{L_1\!\rightarrow L_2}\) transitions, exhibits nearly orthogonal dipole directions, but carries much weaker oscillator strength. The assignment of the doublet to interlayer excitons is therefore supported by the energetic correspondence and the calculated real-space character, while the observed polarization pattern and visibility indicate that the static relaxed-reference-structure picture does not yet fully capture their optical activity.

To explore a possible origin of this optical-activity mismatch, we consider whether low-energy lattice distortions can mix interlayer states with nearby bright intralayer excitons. Phonon calculations reveal several low-energy modes with strong Pb-I buckling character that are predominantly localized on one of the two inorganic layers (see the schematic in Fig.~\ref{fig4}c and the SI). The distortion used here, shown schematically in Fig.~\ref{fig4}c and discussed in detail in the SI, was selected as a representative low-energy symmetry-breaking mode that most clearly illustrates this mixing mechanism. Relative to the relaxed reference structure, it breaks the centrosymmetry of the structure and the near-equivalence of the two inorganic layers and strongly modifies the band-edge shape and alignment (Fig.~\ref{fig4}d,e). A Rashba-type band splitting appears here as a consequence of the local symmetry breaking and shifts the lowest-energy direct transitions slightly away from $\Gamma$. In the present context, this band-level effect is not invoked as the primary origin of the low-energy intralayer fine structure, but as one consequence of the same distortion that also enhances mixing between intra- and interlayer excitonic configurations.

\(G_0W_0\)+BSE calculations for this phonon-displaced structure show much stronger mixing between intra- and interlayer excitonic configurations than in the relaxed reference structure (Fig.~\ref{fig4}f,g). In this case, interlayer states gain oscillator strength and exhibit nearly orthogonal in-plane dipole orientations in the relevant energy range. The assignment of the higher-energy PL doublet to interlayer excitons remains based on the reference-structure energetics, splitting, and real-space character. The distorted-structure calculation therefore illustrates a plausible route by which lattice-coupled mixing can account for the enhanced visibility and altered polarization of the higher-energy doublet in experiment. The assignment of that doublet to interlayer excitons, however, remains based on the reference-structure energetics, splitting, and real-space character.

\section{Conclusions}
In conclusion, polarization-resolved photoluminescence combined with first-principles \(G_0W_0\)+BSE calculations provides an experimentally anchored picture of the excitonic fine structure in \(\mathrm{(PEA)_2PbI_4}\). The main outcome is a symmetry- and layer-resolved assignment of the low-energy intralayer manifold together with the identification of the weak higher-energy doublet as interlayer excitons.
\begin{figure}[ht]
    \centering
    \includegraphics[width=\columnwidth]{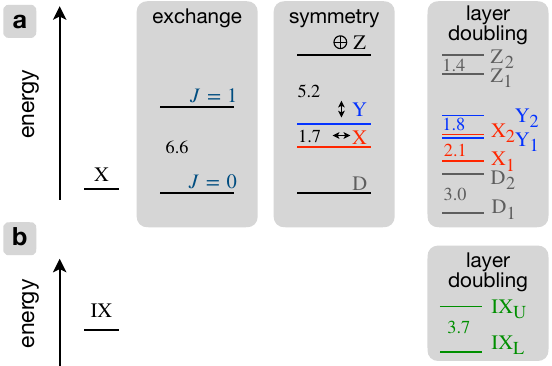}
        \caption{Summary schematic of the experimentally extracted and tentatively inferred excitonic energy scales in \(\mathrm{(PEA)_2PbI_4}\). The dominant bright intralayer transitions and their \(\mathrm{L}_1/\mathrm{L}_2\)-resolved splittings are extracted from polarization-resolved PL, while the positions of the corresponding \(\mathrm{D}\)- and \(\mathrm{Z}\)-like partners are estimated more tentatively from weaker spectral features. The weak higher-energy doublet is included as the assigned interlayer exciton manifold.}
    \label{fig5}
\end{figure}

For the low-energy manifold, the central result is that the observed fine structure can be understood within an intralayer exciton framework that resolves both symmetry and layer degree of freedom. The bright emission multiplet arises from a doubled intralayer manifold associated with the two inequivalent layers of the relaxed structure, and comparative calculations indicate that the corresponding splitting is governed primarily by layer-dependent offsets rather than strong interlayer hybridization. The low-energy PL therefore does not require Rashba splitting or exciton-polaron sidebands as its primary explanation, even though symmetry-breaking lattice distortions may still be relevant for the optical activity of the interlayer excitons.\cite{Do20b,Neutzner18}

The weak higher-energy doublet is assigned to interlayer excitons based on its energetic position, splitting, and calculated real-space character. At the same time, the static reference-structure calculation does not reproduce its observed visibility and nearly orthogonal in-plane polarization axes. Distortion-based calculations suggest that lattice-coupled mixing with nearby bright intralayer excitons provides a plausible explanation for the remaining discrepancy in visibility and polarization.

In the broader context of exciton fine-structure studies in layered lead-halide perovskites,\cite{Blancon20,Do20b,Posmyk22,Baranowski22b,Quarti24} these results support a differentiated view in which electron--hole exchange and in-plane/out-of-plane anisotropy form the established fine-structure background, while layer doubling within the two-layer unit cell, and distinct interlayer excitons emerge here as additional experimentally relevant ingredients of the excitonic spectrum. Figure~\ref{fig5} summarizes this semiquantitative picture by combining the experimentally extracted bright-state splittings with the more tentatively inferred \(\mathrm{D}\)- and \(\mathrm{Z}\)-state positions and the assigned interlayer doublet.

Further progress will require more direct tests of the interlayer-exciton assignment, especially of the \(\mathrm{D}\)- and \(\mathrm{Z}\)-like partner positions and of the origin of the unusual interlayer-exciton optical activity. Resonant and time-resolved spectroscopies should be particularly informative, complemented by temperature-, strain-, or field-dependent measurements. On the theory side, treatments including dynamic lattice fluctuations and exciton-phonon coupling will be needed to determine under which conditions interlayer excitons become bright and acquire the observed polarization pattern.

\begin{acknowledgement}
L.L. and F.L. acknowledge funding by the Dutch Research Council (NWO) through Grant No. VI.Vidi.223.072, EuroHPC for awarding access to the Leonardo Booster supercomputer at CINECA, Italy, and computing resources provided but the Dutch national supercomputing center Snellius supported by the SURF Cooperative.
\end{acknowledgement}

\begin{suppinfo}
The Supporting Information contains
\begin{itemize}
    \item Synthesis protocol and experimental details
    \item Computational details and convergence tests for ground-state, excited-state $G_0W_0$+BSE, and phonon calculations
    \item Additional results (structural parameters, band structures and excited-state energies and oscillator strengths) for relaxed reference structure, Cs$_2$PbI$_4$ model structures, and phonon-displaced structure.
\end{itemize}

\end{suppinfo}

\bibliography{references}


\end{document}